\documentclass[11pt,twoside]{article}

\usepackage{asp2021}

\aspSuppressVolSlug
\resetcounters

\begin{document}

\title{Localization and Confidence Region Estimation of Short GRBs with the COSI BGO Shield Using a HEALPix-Based Deep Learning Approach}


\author{
N.~Parmiggiani,$^{1}$ 
A.~Bulgarelli,$^{1}$ 
G.~Panebianco,$^{1}$ 
E.~Burns,$^{2}$ 
E.~Neights,$^{3,4}$ 
V.~Fioretti,$^{1}$ 
I.~Martinez-Castellanos,$^{5,6}$ 
L.~Castaldini,$^{1}$ 
A.~Ciabattoni,$^{7,1}$ 
A.~Di~Piano,$^{1}$ 
R.~Falco,$^{1}$ 
S.~Gallego,$^{8}$ 
G.~Mustafa,$^{9,1}$ 
P.~Patel,$^{10}$ 
A.~Rizzo,$^{11}$ 
E.~A.~Wulf,$^{12}$ 
D.~H.~Hartmann,$^{13}$ 
C.~A.~Kierans,$^{4}$ 
J.~A.~Tomsick,$^{14}$ and
A.~Zoglauer$^{14}$
}

\affil{$^{1}$INAF/OAS Bologna, Via P. Gobetti 93/3, 40129 Bologna, Italy;
\email{nicolo.parmiggiani@inaf.it}}

\affil{$^{2}$Department of Physics and Astronomy, Louisiana State University, Baton Rouge, LA 70803, USA}

\affil{$^{3}$George Washington University, 2121 I St NW, Washington, DC 20052, USA}

\affil{$^{4}$NASA Goddard Space Flight Center, 8800 Greenbelt Road, Greenbelt, MD 20771, USA}

\affil{$^{5}$Department of Astronomy, University of Maryland, College Park, MD 20742, USA}

\affil{$^{6}$Center for Research and Exploration in Space Science and Technology, NASA/GSFC, Greenbelt, MD 20771, USA}

\affil{$^{7}$Department of Physics and Astronomy ``Augusto Righi'', University of Bologna, Via P. Gobetti 93/2, 40129 Bologna, Italy}

\affil{$^{8}$Institut f\"{u}r Physik \& Exzellenzcluster PRISMA+, Johannes Gutenberg-Universit\"{a}t Mainz, 55099 Mainz, Germany;}

\affil{$^{9}$Department of Computer Science and Engineering, University of Bologna, Via Zamboni 33, 40126 Bologna, Italy}

\affil{$^{10}$George Mason University, 4400 University Dr, Fairfax, VA 22030, USA;}

\affil{$^{11}$INAF/OA Catania, Via Santa Sofia 78, 95123 Catania, Italia}

\affil{$^{12}$U.S. Naval Research Laboratory, 4555 Overlook Ave., SW Washington, DC 20375, USA}

\affil{$^{13}$Department of Physics and Astronomy, Clemson University, Kinard Lab of Physics, Clemson, SC 29634-0978, USA}

\affil{$^{14}$Space Sciences Laboratory, 7 Gauss Way, University of California, Berkeley CA 94720-7450, USA}

\paperauthor{N.~Parmiggiani}{nicolo.parmiggiani@inaf.it}{0000-0002-4535-5329}
{INAF/OAS Bologna}{}
{Bologna}{BO}{40129}{Italy}

\paperauthor{A.~Bulgarelli}{andrea.bulgarelli@inaf.it}{0000-0001-6347-0649}
{INAF/OAS Bologna}{}
{Bologna}{BO}{40129}{Italy}

\paperauthor{G.~Panebianco}{gabriele.panebianco@inaf.it}{0000-0002-3410-8613}
{INAF/OAS Bologna}{}
{Bologna}{BO}{40129}{Italy}

\paperauthor{E.~Burns}{ericburns@lsu.edu}{0000-0002-2942-3379}
{Louisiana State University}{Department of Physics and Astronomy}
{Baton Rouge}{LA}{70803}{USA}

\paperauthor{E.~Neights}{eneights@gwmail.gwu.edu}{0009-0005-0762-4507}
{George Washington University}{}
{Washington}{DC}{20052}{USA}

\paperauthor{V.~Fioretti}{valentina.fioretti@inaf.it}{0000-0002-6082-5384}
{INAF/OAS Bologna}{}
{Bologna}{BO}{40129}{Italy}

\paperauthor{I.~Martinez-Castellanos}{imc@umd.edu}{0000-0002-2471-8696}
{University of Maryland}{Department of Astronomy}
{College Park}{MD}{20742}{USA}

\paperauthor{L.~Castaldini}{luca.castaldini@inaf.it}{0009-0000-5501-4328}
{INAF/OAS Bologna}{}
{Bologna}{BO}{40129}{Italy}

\paperauthor{A.~Ciabattoni}{alex.ciabattoni@inaf.it}{0009-0002-7005-0380}
{University of Bologna}{Department of Physics and Astronomy ``Augusto Righi''}
{Bologna}{BO}{40129}{Italy}

\paperauthor{A.~Di~Piano}{ambra.dipiano@inaf.it}{0000-0002-9894-7491}
{INAF/OAS Bologna}{}
{Bologna}{BO}{40129}{Italy}

\paperauthor{R.~Falco}{luca.castaldini@inaf.it}{0009-0004-1676-7596}
{INAF/OAS Bologna}{}
{Bologna}{BO}{40129}{Italy}

\paperauthor{S.~Gallego}{sgallego@uni-mainz.de}{0000-0002-2664-8804}
{Johannes Gutenberg-Universit\"{a}t Mainz}{Institut f\"{u}r Physik, PRISMA+}
{Mainz}{RP}{55099}{Germany}

\paperauthor{G.~Mustafa}{luca.castaldini@inaf.it}{0009-0005-1088-9119}
{University of Bologna}{Department of Computer Science and Engineering}
{Bologna}{BO}{40126}{Italy}

\paperauthor{P.~Patel}{parshad.patel.civ@us.navy.mil}{0000-0002-8835-8941}
{George Mason University}{}
{Fairfax}{VA}{22030}{USA}

\paperauthor{A.~Rizzo}{alessandro.rizzo@inaf.it}{0009-0003-4341-2988}
{INAF/OA Catania}{}
{Catania}{CT}{95123}{Italy}

\paperauthor{E.~Wulf}{eric.a.wulf.civ@us.navy.mil}{0000-0002-9577-7888}
{U.S. Naval Research Laboratory}{}
{Washington}{DC}{20375}{USA}

\paperauthor{D.~Hartmann}{hdieter@clemson.edu}{0000-0002-8028-0991}
{Clemson University}{Department of Physics and Astronomy}
{Clemson}{SC}{29634}{USA}

\paperauthor{C.~A.~Kierans}{carolyn.a.kierans@nasa.gov}{0000-0001-6677-914X}
{NASA Goddard Space Flight Center}{}
{Greenbelt}{MD}{20771}{USA}

\paperauthor{J.~A.~Tomsick}{jtomsick@berkeley.edu}{0000-0001-5506-9855}
{University of California, Berkeley}{Space Sciences Laboratory}
{Berkeley}{CA}{94720}{USA}

\paperauthor{A.~Zoglauer}{zoglauer@berkeley.edu}{0000-0001-9067-3150}
{University of California, Berkeley}{Space Sciences Laboratory}
{Berkeley}{CA}{94720}{USA}



\begin{abstract}
The Compton Spectrometer and Imager is a NASA satellite mission under development that will survey the entire sky in the 0.2-5 MeV range using a wide-field germanium detector array, surrounded on the sides and bottom by active shields (the Anticoincidence Subsystem, ACS). The ACS aims to suppress and monitor background events, as well as detect transient sources, such as Gamma-Ray Bursts (GRBs), through its onboard triggering algorithm. The data related to GRBs are sent to the ground and analyzed by an automated pipeline to localize the GRBs and share their positions with the community. In this work, we present a brief GRB localization method based on ACS data, utilizing deep learning (DL) techniques, which can estimate the 90\% confidence region, including cases where it is split into multiple areas. To address this, we developed a neural network classifier that predicts the GRB location as a probability distribution across the sky map following the HEALPix framework. The distribution can be used to compute the 90\% confidence regions. Future work will compare this DL-based localization approach with classical methods such as $\chi^2$ fitting and Maximum Likelihood Estimation.
\end{abstract}



\section{Introduction}
The Compton Spectrometer and Imager (COSI, \citet{2024icrc.confE.745T}) is a NASA satellite mission under development that will survey the entire sky in the 0.2-5 MeV range with a wide-field gamma-ray telescope. Its main instrument consists of a germanium detector array, surrounded on the sides and bottom by bismuth germanate (BGO) scintillator active shields (the Anticoincidence Subsystem, ACS). The ACS both suppresses and monitors background events and enables the detection of transient sources. COSI will include an onboard triggering algorithm capable of identifying Gamma-Ray Bursts (GRBs) in the ACS and transmitting data to the ground for further analysis. The GRBs will be localized by an automated pipeline, with the localizations promptly sent to the community. The ACS is composed of 22 BGO crystals, ten on the bottom side and three on each lateral side (left panel of Figure \ref{figure:bgo_scheme}). Each crystal is coupled to a 3x3 array of 6 mm Silicon Photomultipliers (SiPM), each with two gain channels: a high-gain "gamma" channel ($\sim$80 keV - 2 MeV) and a low-gain "proton" channel (>2 MeV). Both channels are read out by an Application-Specific Integrated Circuit (ASIC). Count rates are recorded every 50 ms. The analyses in this work use only the "gamma" channel. The Light Curves of the ACS segments are obtained by grouping a lateral wall per ASIC with two further ASICs reading out the bottom wall. The final two bottom BGO modules are combined with two lateral walls as illustrated in the right panel of Figure \ref{figure:bgo_scheme} to optimize the readout channels. The red boxes indicate the aggregation of bottom modules into lateral walls. \articlefiguretwo{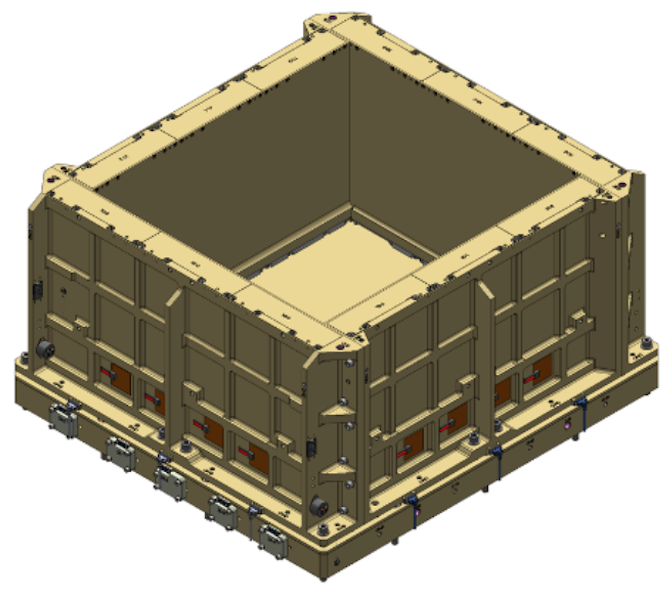}{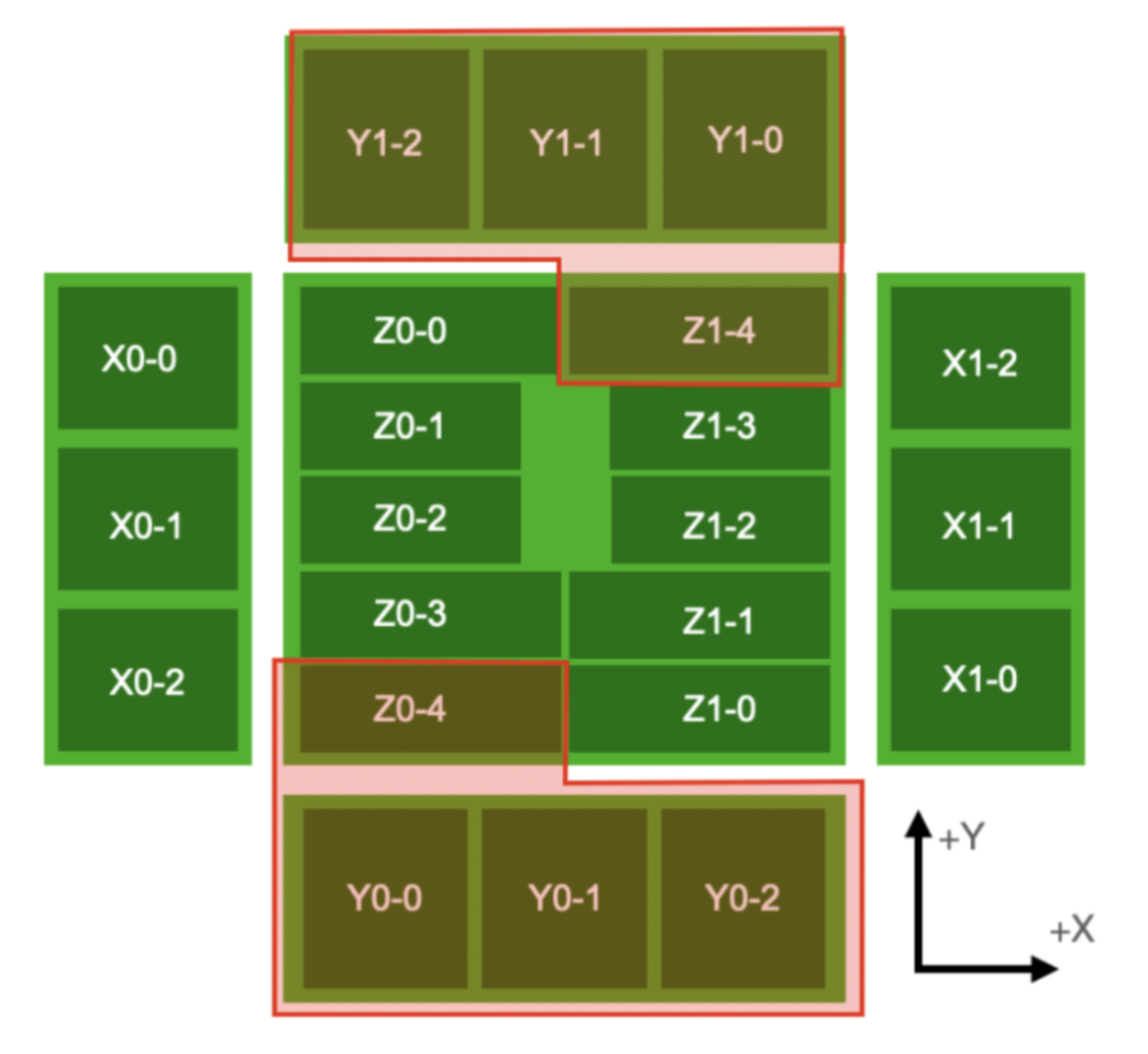}{figure:bgo_scheme}{\emph{Left:} COSI ACS Design. \emph{Right:} BGO crystal scheme.} 

In this work, we present two short GRB (sGRB) localization methods based on ACS data and developed using deep learning (DL) techniques. Different network architectures were evaluated to estimate localization uncertainties at the 90\% confidence level, including cases where the confidence region is split into multiple areas. In previous work \citet{Parmiggiani2026COSI}, we compared the localization performance of our DL model with the $\chi^2$ fit \citep{Connaughton_2015} and the bc-tools localization methods \citep{martinez_castellanos_2025_15419961}. However, that comparison was limited to the angular distance error between the simulated sGRB positions and the localized positions, since the DL model used in that work could not estimate the 90\% confidence error regions. The DL model presented in this manuscript is capable of estimating the 90\% confidence error areas, allowing for a direct comparison with classical localization methods.

\section{Methods}

We simulated two sGRB datasets using MEGAlib \citep{ZOGLAUER2006629} for training and testing purposes. The simulation procedure is described in detail in \citet{Parmiggiani2026COSI}. The testing dataset contains 49,152 sGRBs distributed following the HEALPix coordinate system (nside=64), while the training dataset contains 100,000 sGRBs randomly sampled from the HEALPix positions with nside=128. The sGRBs are simulated with a variable flux and three different spectral models described in \citet{Parmiggiani2026COSI}. 

The first DL model we developed is a probabilistic regressor designed to reconstruct the sGRB arrival direction. The model takes as input the count rates of the ACS panels and outputs a full multivariate normal probability distribution over four correlated directional components ($sin\theta$, $cos\theta$, $sin\phi$, $cos\phi$). The architecture consists of a Feedforward Neural Network with three hidden layers of 256, 128, and 64 neurons, respectively, and the LeakyReLU activation function, similar to the DL model presented in \citet{Parmiggiani2026COSI}. Dropout regularization is applied during training to reduce overfitting. The probabilistic formulation enables the model to provide not only directional estimates but also associated uncertainties. 

The second DL model maps the ACS panel counts to a discrete probability distribution over sky positions represented using the HEALPix framework. The architecture of the network is identical to that of the first model except for the output layer, which uses a softmax activation function to produce a normalized probability distribution over all HEALPix pixels. The labels of the training dataset are one-hot encoded HEALPix maps representing the true source position. To improve training stability, the true positions are smoothed to avoid collapsing onto a single pixel.

\section{Results}

The results show that the probabilistic regression model outputs a multivariate normal distribution, describing sGRB localization uncertainties as elliptical error regions. This approach suits cases where the error distribution is unimodal and Gaussian-like. However, it cannot capture complex or multimodal error structures caused by detector geometry. In contrast, the HEALPix-based probabilistic model, which predicts a probability distribution over the sky map, enables, for the first time in our analysis with DL models, the estimation of multimodal error regions covering multiple possible sky areas, as shown in Figure \ref{fig:results}. The HEALPix probability maps reveal richer spatial structures, capturing several peaks in the predicted localization probability.  \articlefigure[width=0.65\textwidth]{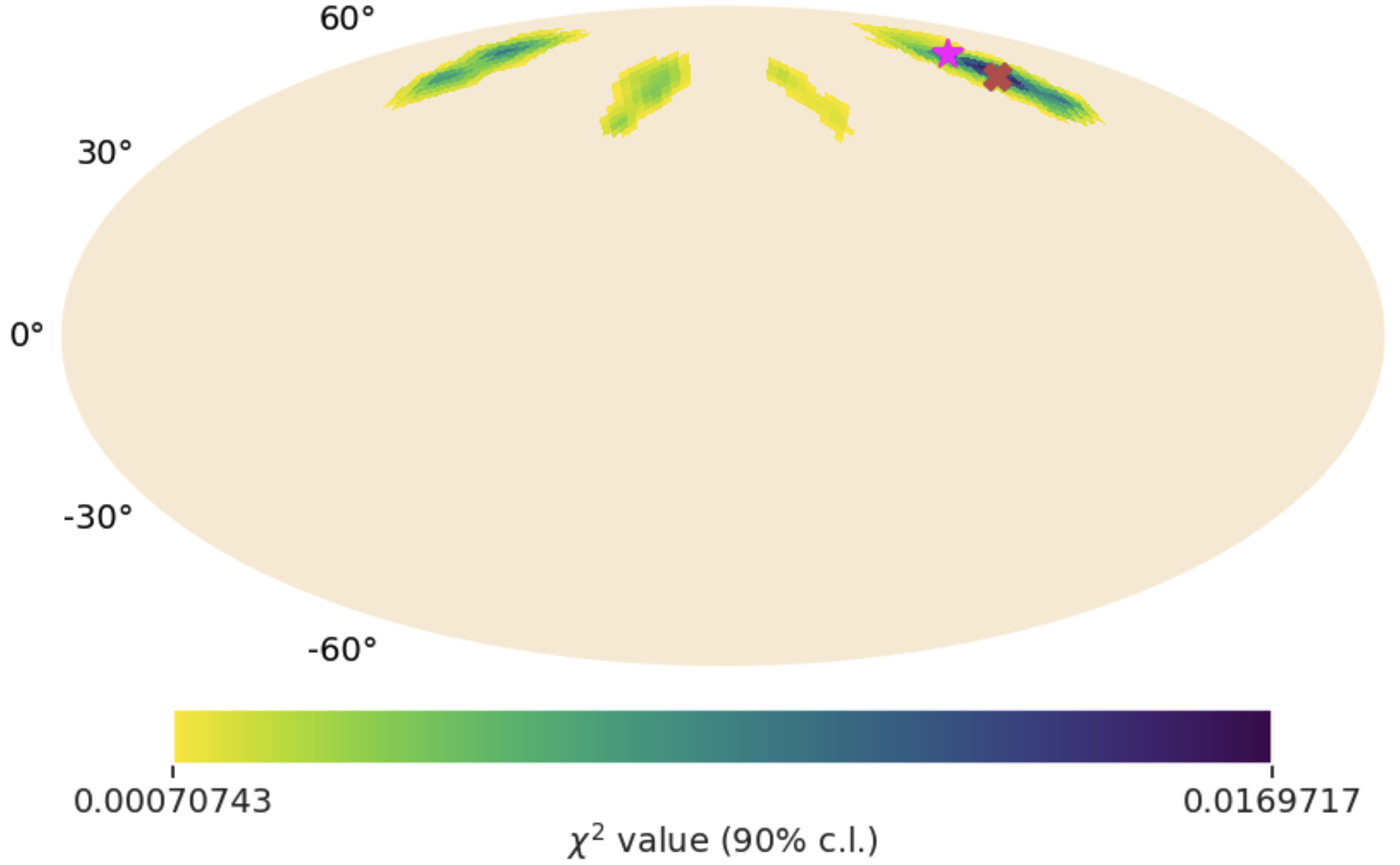}{fig:results}{Examples of 90\% c.l. error regions (997 deg$^2$) with multimodal behaviour obtained with the second DL model. The purple star marker indicates the simulated sGRB positions, while red X markers indicate the sGRB localizations.
}

\section{Conclusion}

The results of this work demonstrate that the HEALPix-based classification model can successfully estimate GRB localization error regions from COSI-ACS data, even when the probability distributions are multimodal and have multiple disjoint regions. This approach extends the model's applicability beyond simple Gaussian assumptions, providing a more realistic representation of localization uncertainty. Ongoing analyses aim to further characterize these error regions and compare them with those obtained using classical methods, such as MLE and $\chi^2$ fitting.

\acknowledgements The Compton Spectrometer and Imager is a NASA Explorer project led by the University of California, Berkeley with funding from NASA under contract 80GSFC21C0059. This work is supported by the Italian Space Agency, partially within contract ASI/INAF No. 2024-11-HH.0.

\bibliography{090}  


\end{document}